\documentclass[twocolumn]{aastex62}

\usepackage{amssymb}
\usepackage{color}
\usepackage{amsmath}
\usepackage{tablefootnote}
\usepackage{comment}
\usepackage{lipsum}
\usepackage{hyperref}
\usepackage{soul}
\usepackage{graphicx}
\usepackage{natbib}
\usepackage{comment}
\usepackage{multirow}
\begin{document}

\title{Using a Neural Network Classifier to Select Galaxies with the Most Accurate Photometric Redshifts}
\author[0000-0002-7767-5044]{Adam Broussard}
\affiliation{Department of Physics and Astronomy, Rutgers, The State University of New Jersey, 136 Frelinghuysen Rd, Piscataway, NJ 08854, USA}
\email{adamcbroussard@physics.rutgers.edu}

\author[0000-0003-1530-8713]{Eric Gawiser}
\affiliation{Department of Physics and Astronomy, Rutgers, The State University of New Jersey, 136 Frelinghuysen Rd, Piscataway, NJ 08854, USA}
\email{gawiser@physics.rutgers.edu}
\date{August 2021, Accepted to ApJ}

\begin{abstract}
The Vera C. Rubin Observatory Legacy Survey of Space and Time (LSST) will produce several billion photometric redshifts (photo-$z$'s), \replaced{making it advantageous for}{enabling} cosmological analyses to select a subset of galaxies \replaced{that have}{with} the most accurate photo-$z$.  \replaced{To this end, we}{We} perform initial redshift fits on Subaru Strategic Program galaxies with deep $grizy$ photometry using Trees for Photo-Z (TPZ) before applying a custom neural network classifier (NNC) tuned to select galaxies with $(z_\mathrm{phot} - z_\mathrm{spec})/(1+z_\mathrm{spec}) < 0.10$.  We \replaced{explore methods of mitigating training sample bias by considering}{consider} four cases of training and test sets ranging from an idealized case to using data augmentation to increase the representation of dim galaxies in the training set.  \deleted{We show that typical photo-$z$ uncertainties can be used to select galaxy subsets with improved photo-$z$ accuracy.  }Selections made using the NNC yield significant further improvements in outlier fraction and photo-$z$ scatter ($\sigma_z$)\replaced{; as}{ over those made with typical photo-$z$ uncertainties.  As} an example, when selecting the best third of the galaxy sample, the NNC achieves a 35\% improvement in outlier rate and a 23\% improvement in $\sigma_z$ compared to using uncertainties from TPZ.  For cosmology and galaxy evolution studies, \replaced{our photo-$z$ pipeline}{this method} can be tuned to retain a particular sample size or to achieve a desired photo-$z$ accuracy; our results show that it \replaced{should be}{is} possible to retain more than a third of an LSST-like galaxy sample while reducing \deleted{the }$\sigma_z$ by a factor of two compared to the full sample, with one-fifth as many photo-$z$ outliers.  \added{For surveys like LSST that are not limited by shot noise, this method enables a larger number of tomographic redshift bins and hence a significant increase in the total signal-to-noise of galaxy angular power spectra.}

\end{abstract}


% \section*{Figures}
% \begin{itemize}
%     \item Color-magnitude plot showing \texttt{HSC Spec}, \texttt{HSC Phot}, and COSMOS
%     \item \textbf{TO DO} Possible diagram to show NN architecture? (Check Murkwood paper)
%     \item Color-magnitude plot showing \texttt{HSC Phot} and constructed training set for COSMOS closest match run
%     \item zspec vs. zphot plots for each of the (4?) runs in Results
%     \item Nfrac vs. Accuracy plots for COSMOS closest match run in Discussion
%     \item Redshift dependent selection plot in Discussion
% \end{itemize}

\section{Introduction}
% \begin{itemize}
%     \item Galaxy photometric redshifts are difficult to measure, but we have large numbers of photometric redshifts compared to spectroscopic redshifts
%     \item Accurate redshifts are needed not only for intrinsic galaxy properties, but also for cosmological measurements because galaxies need to be placed in to tomographic bins in order to measure cosmological parameters
%     \item Neural networks are one of the most flexible forms of machine learning, but often need large amounts of training data.  We make use of them here because of the large HSC spectroscopic catalog available for training
% \end{itemize}

One of the most powerful probes of cosmology is large scale structure, also known as galaxy clustering (e.g., \citealt{Press1974,Davis1985,Cooray2002, Zehavi2005, Desjacques2018}).  Typically, three-dimensional clustering is performed using accurate distance measurements to map galaxy locations.  Distances to far-away galaxies can be inferred from their spectroscopic redshifts, however this is often impractical due to the relative time expense required to attain the necessary spectroscopy as well as the large galaxy samples required to draw out the clustering signal.  Alternatively, photometric measurements can be obtained for a much larger sample of galaxies in a similar amount of time (e.g., \citealt{Loh1986,Zhan2006,Hildebrandt2010}).  This is particularly evident for upcoming surveys such as the Vera C. Rubin Observatory Legacy Survey of Space and Time (LSST), which will obtain quality photometry for several billion galaxies \citep{LSSTScienceBook,LSSTSRD,Ivezic2019}.  However, a photometric analysis comes at the expense of precise radial distances, as photometric redshifts (photo-$z$'s) that are used as a proxy for radial distance are less precise.

Because the uncertainty in radial galaxy distance inferred from photo-$z$'s is many orders of magnitude larger than uncertainties in the transverse directions caused by astrometric errors, it is challenging to directly study the three-dimensional clustering of galaxies.  Instead, galaxies are often grouped into  tomographic redshift bins \citep[e.g.,][]{Zhan2006, Nicola2020} where angular clustering can be studied.  Galaxies can be assigned to an incorrect redshift bin due to photo-$z$ fit uncertainties.  These typically shift galaxies into neighboring bins, but there also exist systematic degeneracies that can alias galaxies between distant bins, creating photo-$z$ ``outliers".  Incorrectly assigned galaxies reduce the signal in their correct bin while adding noise to the angular clustering measurements in the assigned bin.  Some studies have introduced angular correlation function estimators to account for this mixing of galaxies between bins \citep{Awan2020}.  Nonetheless, reducing photo-$z$ outliers decreases contamination between redshift bins, thereby preventing the loss of valuable cosmological information.

Photometric redshift codes can be broadly divided into two categories: template-based and machine learning-based methods.  Template-based photo-$z$ codes typically use a library of either theoretical galaxy spectra or observed galaxy spectra; sometimes the latter have been dimensionally reduced using a method such as Principal Component Analysis.  Filter curves corresponding to the photometry of the observed catalog are used to calculate photometry for each template across a grid of redshifts, and the (redshift,template,normalization) combination with the smallest $\chi^2$ is chosen as the best fit \citep{Ilbert2006, Brammer2008, Arnouts1999}.  One variation on this method is the extension of template fitting to Bayesian statistics with the addition of priors (e.g., Bayesian Photo-Z; \citealt{Benitez1999}), which enables marginalization over template and normalization to produce a posterior photo-$z$ probability distribution function.  Machine learning-based photo-$z$ codes make use of methods such as feed-forward neural networks (ANNz; \citealt{Lahav2012}), convolutional neural networks (NetZ; \citealt{Schuldt2020}) or random forests (Trees for Photo-Z; \citealt{Kind2014, Kind2013}).  While machine learning methods can provide accurate photo-$z$'s without the need to select templates, they require training samples with measured spectroscopic redshifts.  These training sets should be representative of the observational properties of the galaxy samples that the method will be applied to; the methods typically do not extrapolate well to application sets that are not well-represented in feature space by the training set, and the quantification and amelioration of these differences is an area of ongoing study in machine learning and astronomy \citep{Shai2010,Cranmer2019, Malz2021}.  This introduces its own difficulties, as large spectroscopic samples are relatively rare and often cannot span the full extent of galaxy magnitudes of a photometric catalog.

Data augmentation is the process of modifying training data in an effort to improve the generalizability a machine learning model \citep{Shorten2019}.  In the case of image or textual recognition, this can include transformations applied to existing data such as deformations or rotations \citep{Bloice2017}, but also includes the generation of synthetic data for training \citep{Bird2021}.  In this paper, we detail a new process of data augmentation for photometric data that constructs a synthetic training set with a similar color-magnitude distribution to the test set with unknown redshifts using a smaller, non-representative training set with spectroscopic redshifts.

In Section \ref{Sec: Data}, we present the data we use for our analysis.  Section \ref{Sec: Pipeline} details our method of performing initial photo-$z$ fits before applying a neural network classifier to select highly accurate samples of galaxies.  Section \ref{Sec: TrainApp} describes the four approaches we took in designing training and test sets for our pipeline.  Section \ref{Sec: Results} provides a discussion of our most realistic approach with comparisons against a naive reported uncertainty cut.   Section \ref{Sec: Conclusions} concludes.

\begin{figure*}
    \centering
    \includegraphics[width = 0.8\textwidth]{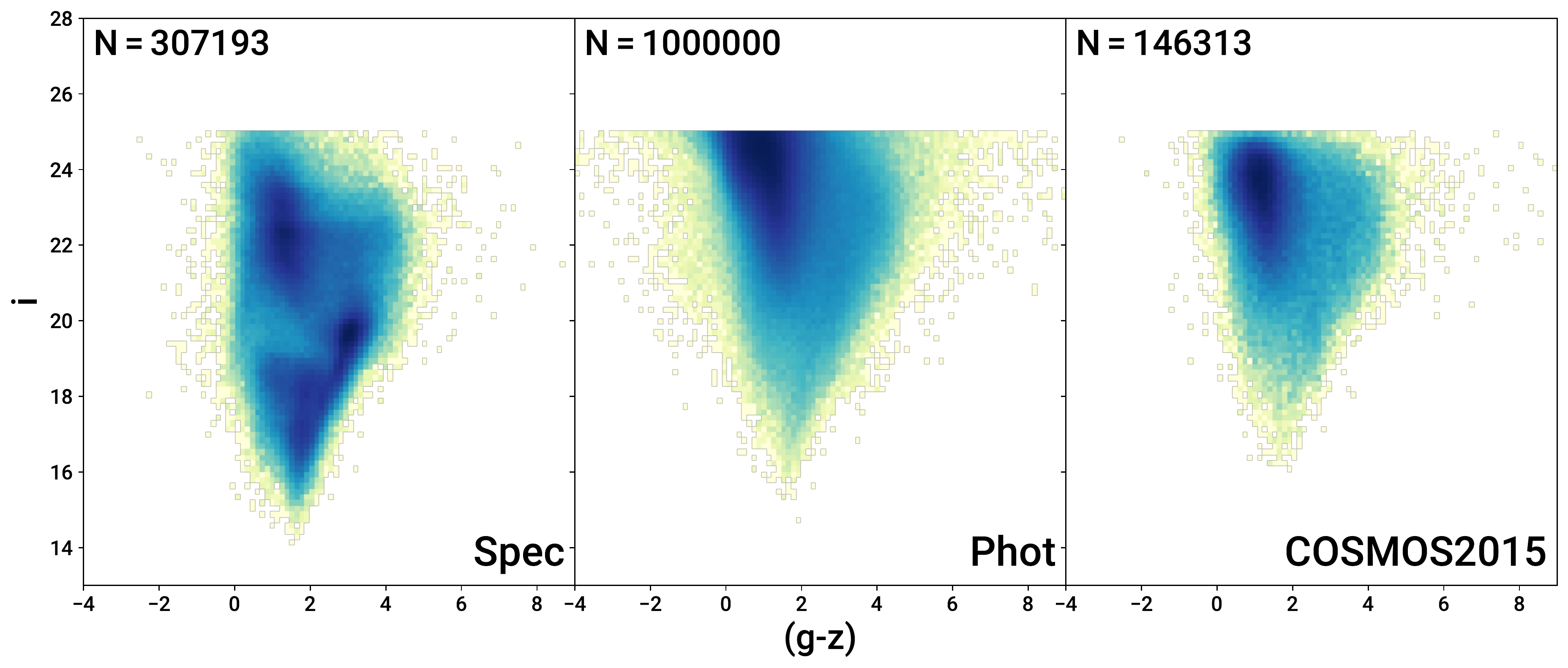}
    \caption{The distribution in $i$-band magnitude vs. $(g-z)$ color for the \texttt{HSC Spec} (left panel), \texttt{HSC Phot} (middle panel), and the \texttt{COSMOS2015} sample formed by matching the \cite{Laigle2016} catalog to \texttt{HSC Phot} (right panel). The primary difference between \texttt{HSC Spec} and the two photometric samples on the right is in completeness for dim galaxies, as the right two panels show robust representation all the way to $i=25$, while the \texttt{HSC Spec} distribution begins to fall off at $i>23$. }
    \label{fig:colormag_phot_spec_cosmos}
\end{figure*}

\section{Data}\label{Sec: Data}

This analysis is modeled on LSST, which will include a large-area $ugrizy$ photometric catalog along with Deep Drilling Field (DDF) photometry over a smaller area offering very high quality photo-$z$'s.  Spectroscopic training sets will be available from DESI, 4MOST, and targeted follow-up, but given the great photometric depth these will inevitably be non-representative.  To mimic the process of estimating photo-$z$'s for early LSST data, we select three analogous data sets to represent each of these components.  Two of these are sourced from the second data release (HSC DC2) of the Hyper Suprime-Cam Subaru Strategic Program (HSC SSP).  The HSC SSP is a photometric galaxy survey that has been awarded 300 nights on the Subaru Telescope starting in March 2014 \citep{Aihara2018}.  DR2 includes data from 174 nights observed at three different depths: \replaced{Wide (300 $\mathrm{deg}^2$ at $i < 26.2$), Deep (\textbf{26} at $i < 27.1$), and UltraDeep (\textbf{4} at $i < 27.7$)}{Wide (900 $\mathrm{deg}^2$ at $i<26.2$) and Deep+UltraDeep (35 $\mathrm{deg}^2$ at $i < 26.7$)} \citep{Aihara2019}.  We focus on the HSC Wide field, which has been observed in five broadband filters (\textit{grizy}) among seven fields.   The HSC SSP Wide catalog contains some 436 million objects processed with hscPipe \citep{Aihara2019}.  Additionally, photometric detections are matched to spectroscopic redshift catalogs from zCOSMOS \citep{Lilly2009}, UDSz \citep{Bradshaw2013, McLure2013}, 3D-HST \citep{Skelton2014,Momcheva2016}, FMOS-COSMOS \citep{Silverman2015,Kashino2019}, VVDS \citep{LeFevre2013}, VIPERS PDR1 \citep{Garilli2014}, SDSS DR12 \citep{Alam2015}, the SDSS IV QSO catalog \citep{Paris2018}, GAMA DR2 \citep{Liske2015}, WiggleZ DR1 \citep{Drinkwater2010}, DEEP2 DR4 \citep{Newman2013}, DEEP3 \citep{Cooper2011, Cooper2012}, and PRIMUS DR1 \citep{Coil2011,Cool2013} along with a homogenized spec-z quality flag for selecting reliably-fit objects. 

\begin{figure}
    \centering
    \includegraphics[width = 0.9\linewidth]{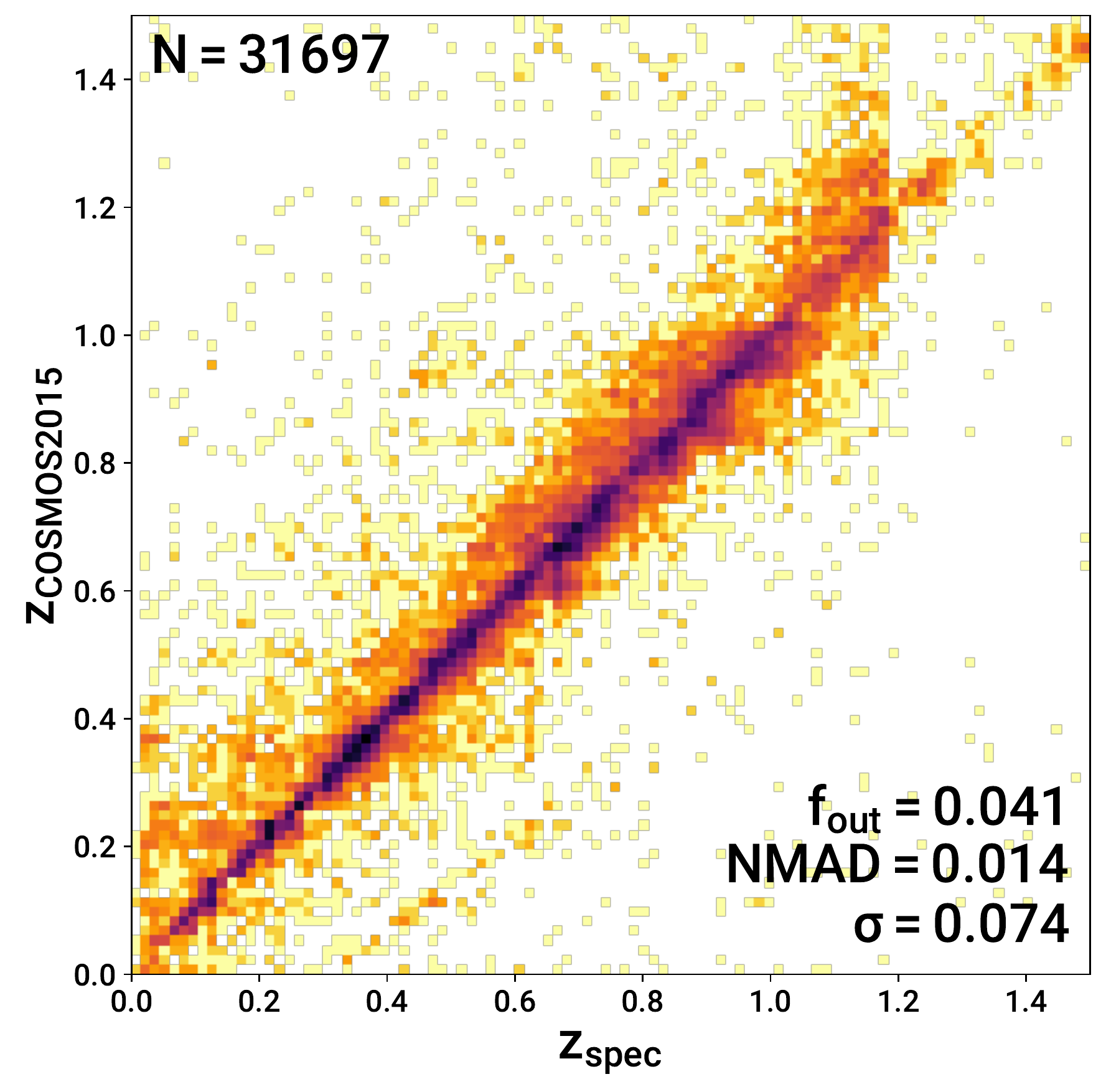}
    \caption{A plot of \texttt{COSMOS2015} 30-band photometric redshift vs. spectroscopic redshift for \texttt{COSMOS2015} galaxies with matching \texttt{HSC Spec} galaxies.  The statistics reported in the lower right corner are for the photo-$z$ error ($\Delta z/(1+z_\mathrm{spec})$).  The small NMAD indicates a tight core distribution of well-fit galaxies, while the larger $\sigma$ indicates the presence of some wide outliers.}
    \label{fig:cosmos2015_zphot_szpec}
\end{figure}

We select two subsets of the HSC Wide catalog that are used to generate training and test sets for the various elements of our machine learning pipeline.  The first, which we refer to as \texttt{HSC Phot} is a random selection of 1 million galaxies, which mimics the depth of the LSST photometric catalog after $\sim2$ years of observations (i.e., $g \le 26.6$, $r \le 26.2$, $i \le 26.2$, $z \le 25.3$, and $y \le 24.5$).  Although the HSC catalogs lack the $u$-band coverage that LSST will have, this implies that LSST will provide somewhat better photo-$z$ performance than we find with HSC photometry. The second data subsample is the set of all galaxies meeting the same magnitude requirements that also have reliable spectroscopic redshifts ($N=307\,193$), which we call \texttt{HSC Spec}.  We consider this to be analogous to the spectroscopic LSST catalog.  

\begin{figure*}
    \centering
    \includegraphics[width = 0.9\textwidth]{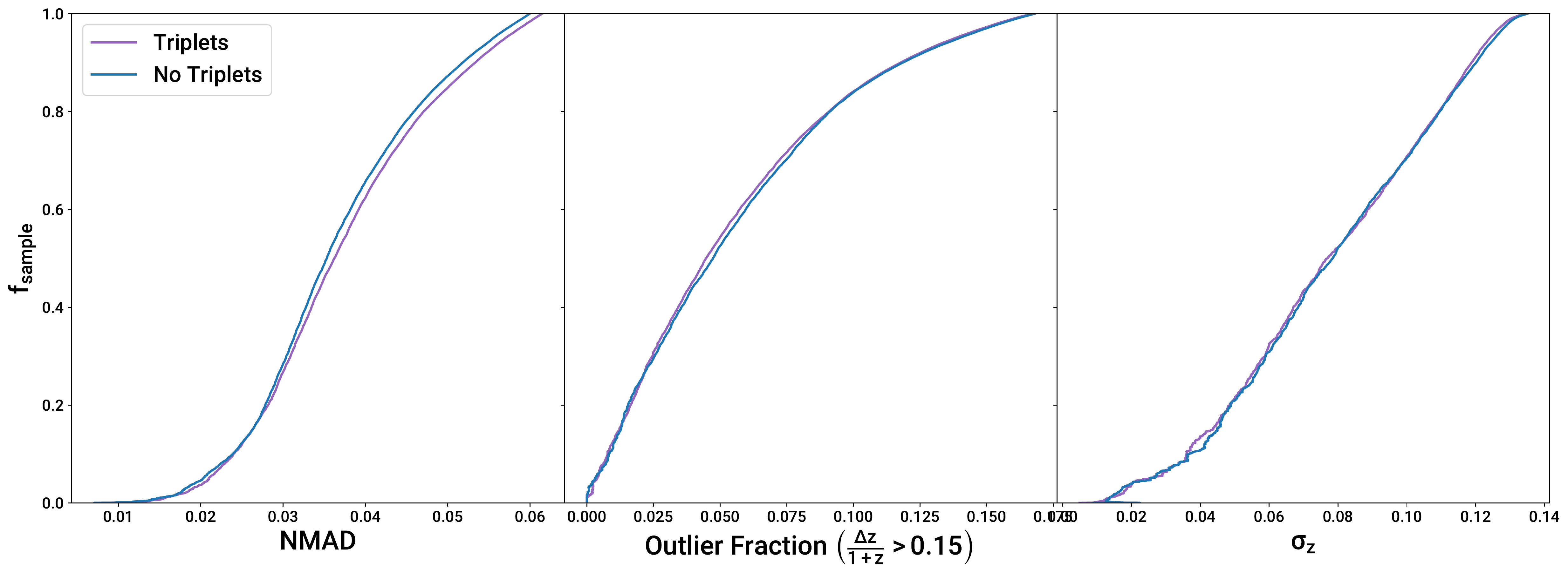}
    \caption{Plots illustrating the tradeoff between retained sample fraction and various metrics when including or excluding band triplets as training features.  The blue curves show results when using band triplets as input features to TPZ and the NNC, while the purple curves show results without.  None of the reported features show a significant difference in performance between the two cases.}
    \label{fig:triplet_comparison}
\end{figure*}

In addition to the HSC SSP catalog, we also make use of the COSMOS2015 catalog produced by \cite{Laigle2016}.  This catalog spans 2 deg$^2$ and adds near-infrared (NIR) and infrared (IR) data from the UltraVISTA and SPLASH surveys to produce a catalog of galaxy SEDs measured in $\sim 30$ broad- and medium-band filters at $\mathrm 1566\AA < \lambda < 108\,335 \AA$.  This catalog has exceptionally well-measured photo-$z$'s for a dimmer sample of galaxies than the \texttt{HSC Spec} sample, making it an ideal test set for our photo-$z$ pipeline.  The full catalog contains 644,327 galaxies with valid redshifts.  This sample narrows to 147,435 galaxies after position matching with HSC photometry to ensure spectral coverage (which we call the \texttt{COSMOS2015} Sample), and again to 32,289 galaxies if we were to sub-select only those galaxies with confirmed spectroscopic redshifts.  For all COSMOS2015 galaxies, we require that $\sigma_\mathrm{COSMOS2015} < 0.1$, $i < 25$, and $\mathrm{NBFILT} >= 25$ to ensure that each object is bright enough to take advantage of COSMOS2015 medium-band coverage to provide precise redshifts.

We show the distribution in $i$-band magnitude vs. $(g-z)$ color for \texttt{HSC Phot}, \texttt{HSC Spec}, and the \texttt{HSC Phot}-matched COSMOS215 sample in Figure \ref{fig:colormag_phot_spec_cosmos}, where the bias towards bright and red galaxies in \texttt{HSC Spec} can be seen as a darkly shaded region along the lower right edge of the distribution in the left panel.  The COSMOS2015 distribution shown in the right panel better matches the magnitude limit of \texttt{HSC Phot}, and has a similar standard deviation in $(g-z)$.  Additionally, we plot photometric vs. spectroscopic redshift for the \texttt{HSC Spec}-matched sub-sample of COSMOS2015 in Figure \ref{fig:cosmos2015_zphot_szpec}.  The small normalized median absolute deviation (NMAD; a measure of distribution width that is not sensitive to outliers) for the photo-$z$ error ($\Delta z/(1+z_{spec})$) indicates that almost all of the \texttt{HSC Spec}-matched COSMOS2015 sample has extremely accurate photo-$z$'s, while the larger $\sigma$ value indicates that those few galaxies without highly accurate fits tend to be wide outliers.  We find that 95\% of galaxies in this sample have $|\Delta z| / (1+z_\mathrm{spec}) < 0.15$; we therefore consider the COSMOS2015 photo-$z$'s to be  analogous to those that will be obtained for galaxies in the LSST DDFs.

\begin{deluxetable*}{c|c|c}
\tablecaption{Pipeline Overview}
% \vspace{-0.2cm}
\tablehead{ \colhead{Stage} & \colhead{Training Features} & \colhead{Output(s)} }
\startdata
~ &$g, r, i, z, y,$&$z_\mathrm{phot},$\\
TPZ &$(g-r), (r-i), (i-z), (z-y),$& $\sigma_\mathrm{TPZ}$\\
~ & $[(g-r)-(r-i)], [(r-i)-(i-z)], [(i-z)-(z-y)]$&$zConf$\\\hline
~ &$z_\mathrm{TPZ}, \sigma_\mathrm{TPZ}, zConf$& \multirow{3}{*}{$C_\mathrm{NNC}$}\\
NNC &$i, (g-r), (r-i), (i-z), (z-y),$&\\
~ & $[(g-r)-(r-i)], [(r-i)-(i-z)], [(i-z)-(z-y)]$&\\
\enddata
\tablecomments{\label{Tbl:training_features} Features used to train each stage of the photo-$z$ pipeline, along with their respective outputs.}
\end{deluxetable*}

\section{Pipeline Design}\label{Sec: Pipeline}

\subsection{Photo-z from TPZ}

We start by performing an initial photometric redshift fit, which we will later use as a baseline for a comparison against the improvements of our pipeline as well as an input to the pipeline itself.  In this work, we use Trees for Photo-Z (TPZ; \citealt{Kind2013}) for this step, but the particular choice of photo-$z$ code is not critical, as we found our pipeline to improve sample redshift accuracy for multiple codes without significant modification (see Appendix \ref{apdx:bpz} for similar results using BPZ, a template-based photo-$z$ code).  TPZ is a machine learning photo-$z$ fitting code that uses random forests to fit photometric redshifts.  TPZ can be used in one of two modes: classification or regression.  The classification mode can be used to perform regression-like photo-$z$ fitting by dividing up the redshift space into many narrow redshift bins, while the regression mode is used to make a point estimate of a galaxy's photo-$z$.  In this work, we make use of the regression mode, as \cite{Kind2013} finds this mode to provide more accurate redshift and error estimates, and we choose to implement 100 trees for the random forest.  Hyperparameters for TPZ include the minimum leaf size as well as $m_*$, the number of features to randomly sub-select for consideration when calculating how to split a given tree node.  We set the minimum leaf size to 30 and $m_*$ to 6, meaning that half of the features are considered for splitting each node.  We limit the TPZ photo-$z$ fits to $0<z_\mathrm{phot}<1.5$, as this approximates the range over which the $4000\,\mathrm{AA}$ break is present in $grizy$ filters, yielding comparatively high-quality photo-$z$'s.

TPZ produces three parameters for each fit galaxy: $z_\mathrm{phot}$, $\sigma_\mathrm{TPZ}$, and $zConf$.  Here, $z_\mathrm{phot}$ is a point estimate of the galaxy's redshift calculated from the mean of the PDF, $\sigma_\mathrm{TPZ}$ is the associated Gaussian uncertainty, and $zConf$, defined as the integrated probability between $z_{phot} \pm \sigma_\mathrm{TPZ}(1+z_{phot})$.

\begin{figure*}
    \centering
    \includegraphics[width = 0.9\textwidth]{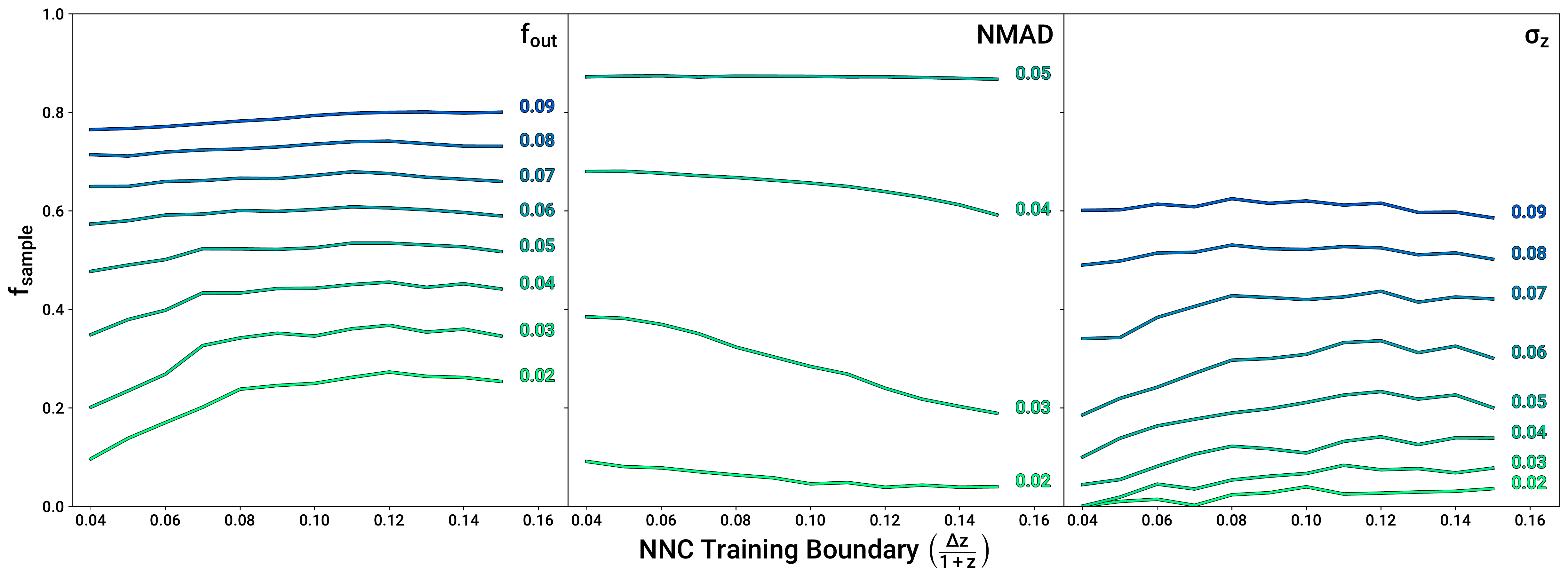}
    \caption{Plots of the sample fraction retained as a function of NNC training boundary for various choices of outlier fraction, NMAD, and $\sigma_z$.  \replaced{The outlier fraction and $\sigma_z$ plot reach a maximum at $|\Delta z|/(1+z) \sim 0.12$, however the NMAD monotonically increases with increasing NNC training boundary}{For horizontal lines corresponding to a fixed sample fraction retained, the outlier fraction and $\sigma_z$ reach minimal values at an NNC training boundary of 0.12 while the NMAD increases}.  As a result, we choose $|\Delta z|/(1+z) = 0.10$ as our NNC training boundary to balance these two effects.}
    \label{fig:acc_var}
\end{figure*}

\subsection{Band Triplets as Features for Training}\label{Sec: bandtriplets}

Both photometric magnitudes and colors are commonly used as training features for machine learning-based photo-$z$ methods. The features we use to train TPZ and our neural network classifier (NNC) afterburner include all available band magnitudes (\textit{grizy}) and all colors between neighboring bands ($(g-r), (r-i), (i-z), (z-y)$). 
  Additionally, we create a new set of spectral features for training that we call \textit{band triplets}, defined as the difference of two neighboring colors e.g., $(g-r)-(r-i)$.  Analogous to the way that colors trace the derivative of the underlying SED, band triplets can be used as a tracer of the SED's second derivative.
  It is possible to construct three band triplets in this way from the HSC $grizy$ bands. 

We completed fits using the training and test sets described in \S~\ref{Sec: MatchCOSMOS}  both with and without band triplets to evaluate their efficacy in outlier fraction (defined as $\Delta z / (1+z) > 0.15$), Normalized Median Absolute Deviation (NMAD), and standard deviation of photo-$z$ error ($\Delta z/(1+z)$) $\sigma_z$.  We show the results in Figure \ref{fig:triplet_comparison}, which indicates that fit accuracy across all shown metrics with band triplets as a feature is nearly identical to fit accuracy without.  This is presumably because the pipeline is able to reproduce such relationships in the course of training, either by approximation (for the random forest used by TPZ)  or by direct linear combination in the case of the neural network classifier (see \S~\ref{Sec:NN}).    Because including triplets does not greatly increase computation time, we retain band triplets as features for each of the approaches detailed in Section \ref{Sec: TrainApp}.

\begin{deluxetable*}{c|c|c}
\tablecaption{Training and Test Set Summary}
% \vspace{-0.2cm}
\tablehead{ \colhead{Approach} & \colhead{Training Set} & \colhead{Test Set} }
\startdata
\textbf{Spec $\rightarrow$ Spec} & \texttt{HSC Spec} & \texttt{HSC Spec} \\\hline
\textbf{Spec $\rightarrow$ COSMOS2015} & \texttt{HSC Spec} & \texttt{COSMOS2015} \\\hline
\textbf{COSMOS2015 $\rightarrow$ COSMOS2015} & \texttt{COSMOS2015} & \texttt{COSMOS2015} \\\hline
\textbf{Match $\rightarrow$ COSMOS2015} & Constructed from \texttt{HSC Spec} and \texttt{COSMOS2015} & \texttt{COSMOS2015}\\
\enddata
\tablecomments{\label{Tbl:train_app} A summary of the various training and test sets used for each successive approach.  Cases where the training and test sets were drawn from the same sample use an 80\%/20\% split.  For a more detailed discussion, see Sections \ref{Sec: SpecSpec}, \ref{Sec: SpecCOSMOS}, \ref{Sec: COSMOSCOSMOS}, and \ref{Sec: MatchCOSMOS}.}
\end{deluxetable*}

\subsection{Neural Network Architecture and Training}
\label{Sec:NN} 

Although artificial neural networks (ANNs) can be traced back to the 1940's \citep{Fitch1944}, they have greatly increased in popularity in the last several decades due to improvements in computational capacity.  Deep neural networks, composed of an input layer, an output layer, and one or more hidden layers, were introduced in 2006 \citep{Hinton2006,Benigo2007, Lecun2015}.  Deep learning is capable of extracting high-level information from complex raw input data by abstracting it multiple times with each hidden layer \citep{Heaton2016}.  According to the universal approximation theorem, any continuous function can be approximated by a feed forward neural network with a single hidden layer \citep{Hornik1989}, however deep neural networks are often better approximators of highly non-linear functions (such as photo-$z$ estimation) than shallow neural networks \citep{Mhaskar2016}.

Our neural network classifier (NNC) consists of four hidden layers, each of which is fully connected.  The layers are comprised of [100, 200, 100, 50] neurons that make use of the Scaled Exponential Linear Unit (SELU) activation functions.  The SELU activation function is implemented along with a LeCun Normal weight initialization to produce a self-normalizing neural network.  Self-normalizing neural networks avoid the the vanishing/exploding gradient problem \citep{Pascanu2012,Haber2018} and train more efficiently than neural nets using activation functions such as Rectified Linear Units \citep{Klambauer2017}.  The output neuron of the NNC uses a sigmoid function with a binary cross-entropy loss function so that it produces values from 0 to 1, indicating redshift accuracy confidence.  We make use of the Adam optimizer \citep{Kingma2014} and implement early stopping with a tolerance of 25 epochs that halts training when the fit accuracy of validation data begins to decrease (and reverts to the best-fit state).

Following the initial TPZ photo-$z$ fit, each of the output values ($z_\mathrm{phot}$, $zConf$, $\sigma_\mathrm{TPZ}$) are passed to the NNC as training features in addition to the original TPZ training features.  The NNC then outputs $C_\mathrm{NNC}$, which has a value between 0 and 1 and identifies the confidence with which each object's photometric redshift has been classified as high-quality (with values close to 0 instead implying an outlier).  As part of this process, we must choose a boundary in $|\Delta z|/(1+z)$ \added{that delineates high quality photo-z's from others.  This initial classification is then used in training the NNC and thus is directly related to the quality of the resulting classifications}.  We elaborate on this choice in the following section (See \S~\ref{Sec: boundary}.  A summary of the features and outputs produced by each of these steps is provided in Table \ref{Tbl:training_features}.

\subsection{Choosing the NNC Training Boundary}\label{Sec: boundary}

To optimize the NNC training boundary, we adopted the training and test sets described in \S~\ref{Sec: MatchCOSMOS} 
and relabeled the training set classes for each possible value of the training boundary.  
Thus, we trained the NNC to classify galaxies using training boundaries of $|\Delta z|/(1+z_\mathrm{spec}) < [0.04, 0.15]$ in increments of 0.01, leading to the results shown in Figure \ref{fig:acc_var}.  The left and right panels indicate that a NNC training boundary of 0.12 would optimize the outlier fraction and $\sigma_z$, however there is a small monotonic increase in NMAD as the NNC training boundary increases shown in the middle panel.  This illustrates a trade-off when training the NNC between tightening the core of the redshift error distribution and reducing outliers.  Therefore we choose a NNC training boundary of 0.10 in an attempt to balance the improvements to $f_\mathrm{out}$ and $\sigma_z$ against the increase in NMAD.  This choice of training boundary results in between 73\% and 92\% of the training set in the positive class, depending on the data used (See \S~\ref{Sec: TrainApp}).

\section{Results} \label{Sec: TrainApp}

\begin{figure}
    \centering
    \includegraphics[width = \linewidth]{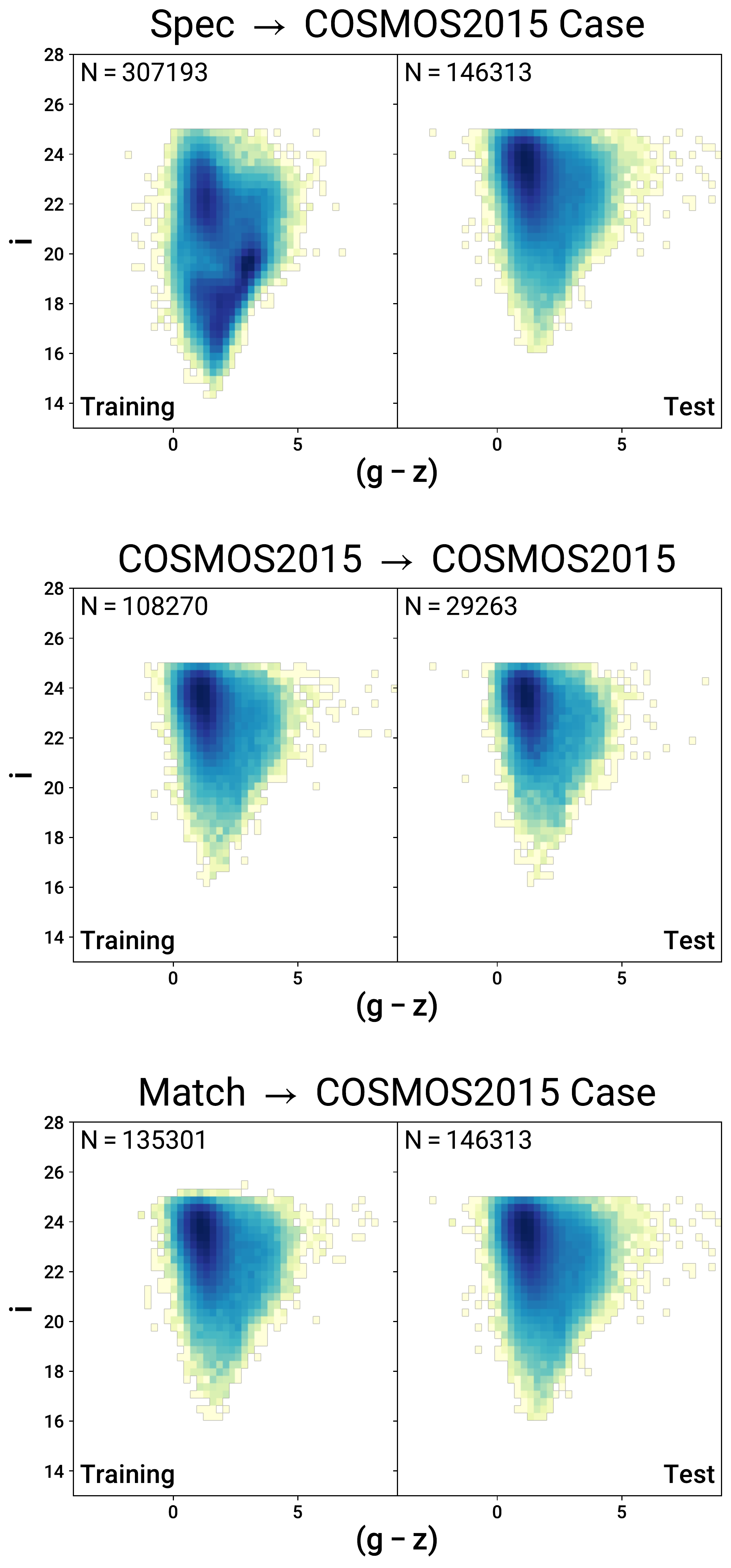}
    \caption{A plot of $i$-band magnitude vs. $(g-z)$ color for the training (left column) and test samples (right column) of each approach detailed in Section \ref{Sec: TrainApp}, excluding the Spec $\rightarrow$ Spec Case (Section \ref{Sec: SpecSpec}), as both the training and test sets are drawn from \texttt{HSC Spec}, and the corresponding color-magnitude distribution is already shown in Figure \ref{fig:colormag_phot_spec_cosmos}.  For details on the construction of each training set-test set pair shown, see Sections \ref{Sec: SpecCOSMOS}, \ref{Sec: COSMOSCOSMOS}, and \ref{Sec: MatchCOSMOS}. }
    \label{fig:all_colormag}
\end{figure}

Although one can naively train on the \texttt{HSC Spec} sample before fitting the \texttt{HSC Phot} sample as an application set, this is unlikely to yield good fits due to the differences in distributions of colors and magnitudes between the two samples.  Indeed, it would be difficult to gauge the true accuracy of the photo-$z$ pipeline in that case, as spectroscopic redshifts are not generally available for the \texttt{HSC Phot} sample.  With this in mind, we designed a series of trial cases for the photo-$z$ pipeline such that each has a unique pair of training and test sets.  The chosen training and test sets, some of which are produced using data augmentation, are designed to reduce the training-test set mismatch and to thereby increase our accuracy in estimating the performance of the model on \texttt{HSC Phot}.  We show color-magnitude distributions for the training and test sets of each approach in Figure~\ref{fig:all_colormag}, summarize them in Table~\ref{Tbl:train_app}, and offer detailed descriptions of each below.

    \subsection{The Spec $\rightarrow$ Spec Case}\label{Sec: SpecSpec}
\begin{figure*}[p]
    \centering
    \includegraphics[width = 0.9\textwidth]{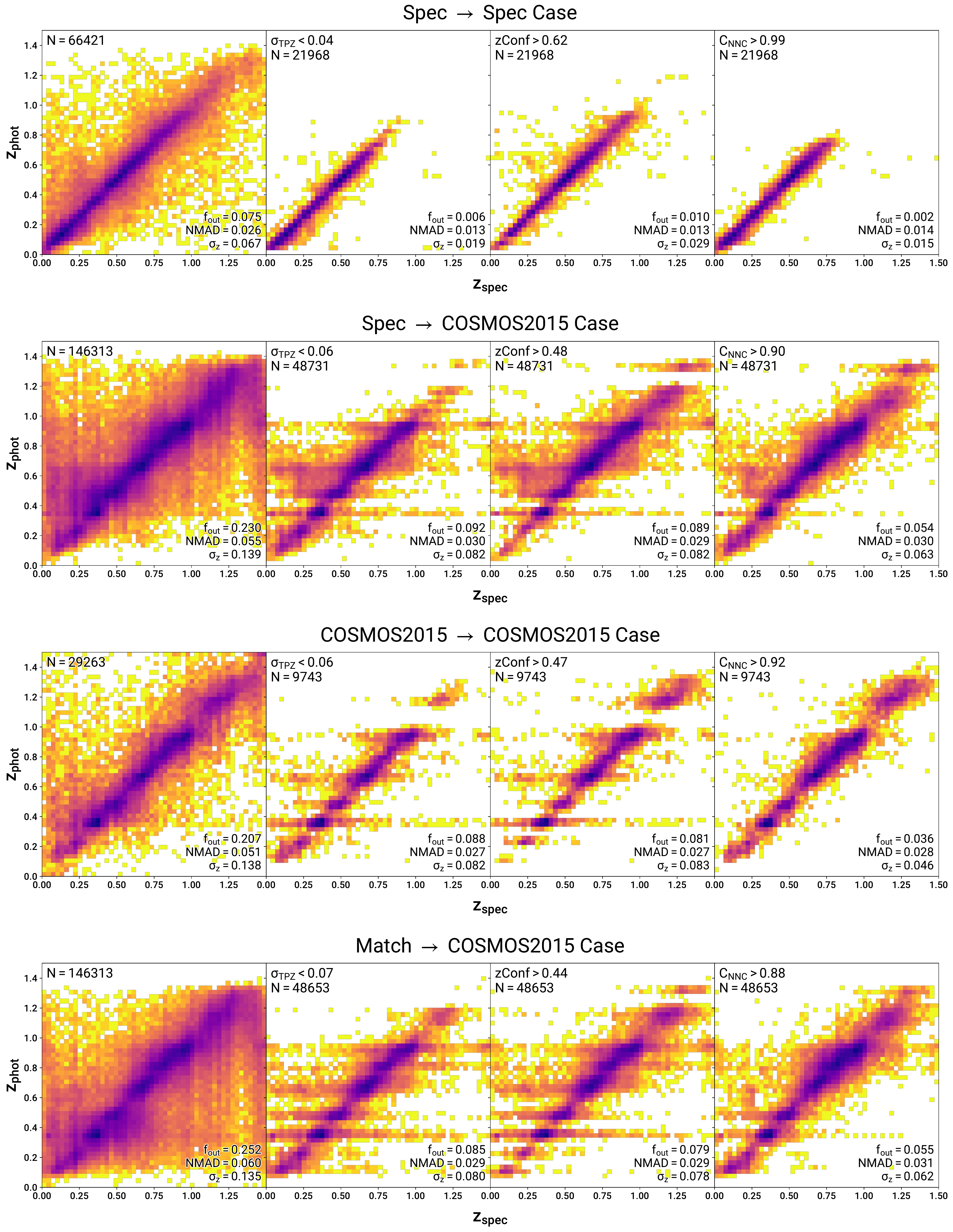}
    \caption{Results in $z_\mathrm{phot}$ vs. $z_\mathrm{spec}$ for each of the approaches detailed in Section \ref{Sec: TrainApp}.  For each approach, the first panel shows the initial photo-$z$ fits performed by TPZ.  The second, third, and fourth panel in each row shows the cuts made in $\sigma_\mathrm{TPZ}$, $zConf$, and $C_\mathrm{NNC}$ that retain the best third of the original sample.  Statistics such as outlier rate ($f_\mathrm{out}$; fraction of objects with $|\Delta z|/(1 + z_\mathrm{spec}) < 0.15$), NMAD, and $\sigma_{z}$ are shown in the lower right corner, while the specific cuts performed in each panel and resulting sample sizes are shown in the upper left.  In general, we find that the NNC-selected sample shown in the fourth panel provides little to no improvement in NMAD compared to the $\sigma_\mathrm{TPZ}$ and $zConf$ cuts, but often improves $f_\mathrm{out}$ by 30\%-40\% and $\sigma_z$ by 35\%-60\%.}
    \label{fig:all_results}
\end{figure*}

    \begin{itemize}
        \item \textit{Training Sample:} 80\% of HSC galaxies with spectroscopic redshifts (\texttt{HSC Spec})
        \item \textit{Test Sample:} Remaining 20\% of HSC galaxies with spectroscopic redshifts (\texttt{HSC Spec})
    \end{itemize}
    While this approach is unrealistic in that the training and test sets both originate from \texttt{HSC Spec}, this serves as an idealized case.  The results provide a best-case performance for TPZ because the training and test sets are both drawn from \texttt{HSC Spec}, and we are able to evaluate our pipeline's performance on ideal photo-$z$'s.  The first row of Figure \ref{fig:all_results} shows plots of $z_\mathrm{spec}$ vs. $z_\mathrm{phot}$ for this case. The first panel shows the results for the full mock test sample while the second, third, and fourth panels show results when using a particular metric to select the best third of the sample (corresponding to $\sigma_\mathrm{TPZ} < 0.04$ in the second panel, $zConf > 0.62$ in the third panel, and $C_\mathrm{NNC}>0.99$ in the fourth).  This selection of one third of the sample is meant to be representative of a cut made on LSST data, but the precise sample fraction retained is variable and will change the resulting photo-$z$ statistics.  The reported statistics in each panel match those of Figure \ref{fig:triplet_comparison}.  Despite retaining 33\% of the full mock sample, all three methods of sample selection offer a significant improvement in terms of outlier rate, $\sigma_z$, and NMAD, with the NNC selection yielding the best results.  In this idealized case  where the training and test set distributions in feature space are identical, the NNC does not improve upon TPZ sample selection.

    \subsection{The Spec $\rightarrow$ COSMOS2015 Case}\label{Sec: SpecCOSMOS}
    \begin{itemize}
        \item \textit{Training Sample:} HSC galaxies with spectroscopic redshifts (\texttt{HSC Spec})
        \item \textit{Test Sample:} Photometry from \texttt{HSC Phot} objects that have counterparts in the COSMOS2015 catalog (the \texttt{COSMOS2015} sample) along with their COSMOS2015 photo-$z$'s as truth.
    \end{itemize}

    While the previous case gives the best-case expected model performance on a test set after training on a fully representative training set, this case approximates the mismatch between training and test sets that could be expected when training on \texttt{HSC Spec} and testing on \texttt{HSC Phot} because it uses the \texttt{COSMOS2015} sample as a test set.  The COSMOS2015 catalog \citep{Laigle2016} contains $\sim30$ photometric bands, and consequently highly accurate photometric redshifts for $i<25$ galaxies, which are bright enough to trigger the medium-bands.  We find that the COSMOS2015 photo-$z$'s for galaxies with $i<25$ have an NMAD of 0.014 compared to our $grizy$ HSC photo-$z$, allowing the construction of a test set where we treat these extremely accurate 30-band photo-$z$'s of COSMOS2015 as truth against which we can test our 5-band photo-$z$ pipeline.  We position-match the COSMOS2015 catalog with $i<25$ to the full HSC catalog of photometric-depth objects and assign the COSMOS2015 30-band photo-$z$'s to the HSC objects as ``true redshifts'' for a test set.
    A comparison of $i$-band magnitude vs. $(g-z)$ color between \texttt{HSC Spec} (left panel) and the constructed test set (right panel) is shown in Figure \ref{fig:all_colormag} in the first row.  The color-magnitude distribution where the training set shows a clear bias towards red and bright galaxies in comparison to the test set.  We show the results of the photo-$z$ fits and subsequent selections in the second row of Figure \ref{fig:all_results}.  Again, we find that each of the selection methods greatly improves the photo-$z$ quality of the resulting sample.  Here, it is evident that the NNC selection shown in the fourth panel is an even greater improvement over $\sigma_\mathrm{TPZ}$ and $zConf$ than in the first case, with $f_\mathrm{out}$ and $\sigma_\mathrm{TPZ}$ decreasing by 59\% and 44\% respectively.  The sample selected by the NNC retains a similar core distribution width as indicated by the NMAD.

\begin{figure*}
    \centering
    \includegraphics[width = \textwidth]{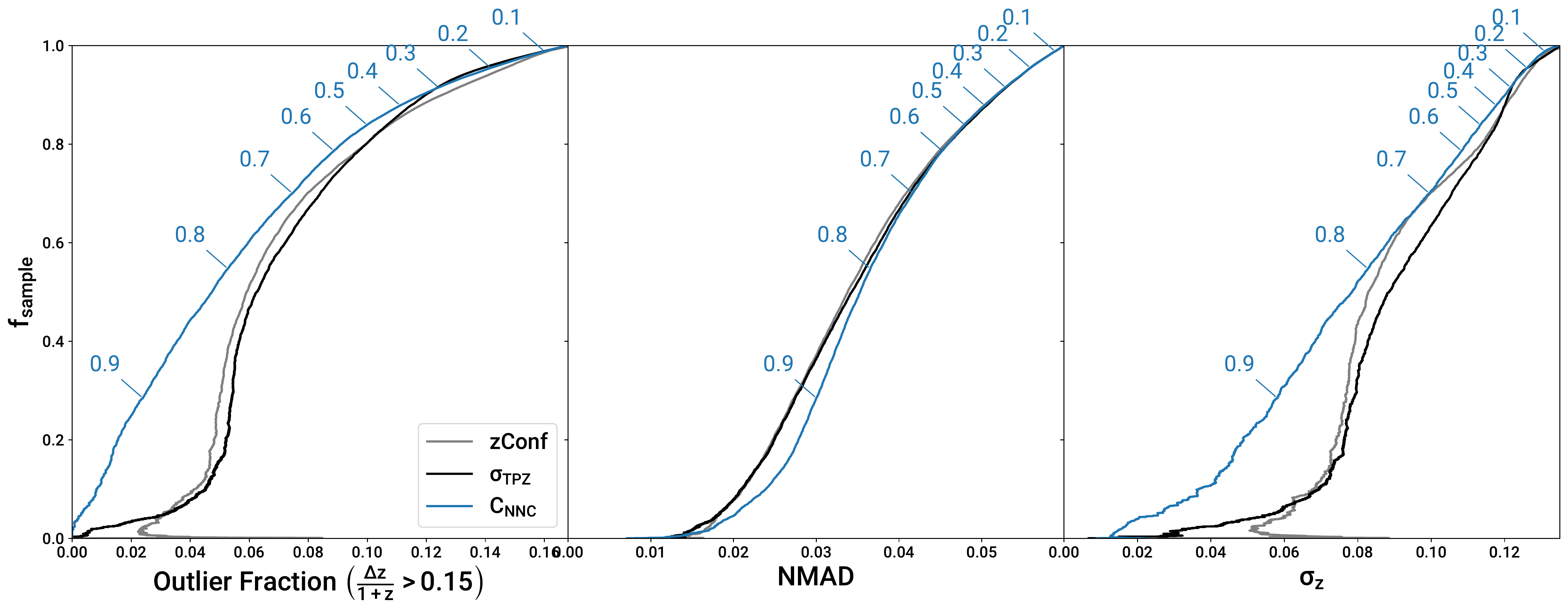}
    \caption{Plots of retained sample fraction ($f_{sample}$) as a function of selected sample accuracy for three metrics: outlier rate (defined as sample fraction with $|\Delta z| / (1+z) > 0.15$), NMAD, and $\sigma_z$.  Labeled points on the blue curves correspond to boundaries in $C_\mathrm{NNC}$.  For comparison, we show curves for cuts in reported photo-$z$ uncertainty ($\sigma_\mathrm{TPZ}$; black), and $zConf$ (a measure of photo-$z$ PDF Gaussianity).  The NNC performs similarly to TPZ on the core of the error distribution, as evidenced by the similar curves in the center panel, but it outperforms both reported TPZ metrics at reducing the outlier rate and $\sigma_z$ in the other two panels.}
    \label{fig:nfrac}
\end{figure*}

    \subsection{The COSMOS2015 $\rightarrow$ COSMOS2015 Case}\label{Sec: COSMOSCOSMOS}

    \begin{itemize}
        \item \textit{Training Sample:} Photometry from 80\% of \texttt{HSC Phot} objects that have counterparts in the COSMOS2015 catalog (the \texttt{COSMOS2015} sample) along with their COSMOS2015 photo-$z$'s.
        \item \textit{Test Sample:}  Photometry from the remaining 20\% of \texttt{HSC Phot} objects that have counterparts in the COSMOS2015 catalog (the \texttt{COSMOS2015} sample) along with their COSMOS2015 photo-$z$'s as truth.
    \end{itemize}
    
    The \texttt{COSMOS2015} sample provides galaxies with a similar distribution of brightnesses and colors to \texttt{HSC Phot} along with high quality (though imperfect) photo-$z$'s.  However, because the \texttt{COSMOS2015} sample is so similar to the intended application set (\texttt{HSC Phot}) but lacks spectroscopic redshifts, we include this case of splitting \texttt{COSMOS2015} into training and test sets as an example bracketing the best-case model performance when using it as a test set.  The third row of Figure \ref{fig:all_results} shows the results of this approach in a similar format to the first two cases.  The first panel shows an initial photo-$z$ fit quality more in line with that of the Spec $\rightarrow$ Spec Case (\S~\ref{Sec: SpecSpec})\added{ as, in both of these cases, the training set is fully representative of the test set}.  The selections in $\sigma_\mathrm{TPZ}$, $zConf$, and $C_\mathrm{NNC}$ are are broader in $|\Delta z| / (1+z)$, as is expected when using photo-$z$'s rather than spec-z's for training.  There is a dearth of galaxies at $1.0 \lesssim z \lesssim 1.2$ in the initial photo-$z$ fits that are nearly completely removed in the $\sigma_\mathrm{TPZ}$ and zConf selections.  By contrast, the NNC selection is able to retain a larger fraction of galaxies in this redshift range while also out-performing the other sample selection techniques in outlier fraction, and $\sigma_z$ by 60\% and 45\% respectively.

    \subsection{The Match $\rightarrow$ COSMOS2015 Case}\label{Sec: MatchCOSMOS}
    \begin{itemize}
        \item \textit{Training Sample:} For each galaxy in the \texttt{COSMOS2015} sample, we find the \texttt{HSC Spec} galaxy that is closest in SED shape.  The photometry of the matching \texttt{HSC Spec} galaxy is then re-scaled to match the normalization of the \texttt{COSMOS2015} galaxy's photometry in the HSC catalog.  We add noise to the rescaled SED to match the HSC catalog uncertainties of the \texttt{COSMOS2015} galaxy's SED.  Throughout this process, we preserve the spec-z of the match galaxy.  In this way, we are able to produce a mock photometric-depth galaxy catalog that includes spec-z's to gauge the accuracy of our photo-$z$ selection.
        \item \textit{Test Sample:} Photometry from \texttt{HSC Phot} objects that have counterparts in the COSMOS2015 catalog (the \texttt{COSMOS2015} sample) along with their COSMOS2015 photo-$z$'s as truth.
    \end{itemize}

    This approach takes advantage of the physical expectation that galaxy SEDs are scalable i.e., that preserving the SED shape preserves the redshift.  We generate the test set by considering each object in \texttt{COSMOS2015} as a normalized 5-dimensional vector in flux space.  We then find the normalized flux vector from the \texttt{HSC Spec} sample that produces the largest dot product as our matched SED.  The matched SED is renormalized while preserving signal-to-noise ratio (SNR) to match the normalization of the corresponding \texttt{COSMOS2015} SED.  Finally, Gaussian noise is added to any bands where the SNR is higher for the matched SED using $\sigma_\mathrm{added} = \sqrt{\sigma_\texttt{HSC Phot}^2 - \sigma_\mathrm{match}^2}$.  We show the color-magnitude distribution of this constructed training set in the last row of Figure \ref{fig:all_colormag} in the left panel, along with that of the \texttt{HSC Phot}-matched \texttt{COSMOS2015} sample in the right panel as the test set.  The training and test sets have similarly shaped color-magnitude distributions despite the training set being comprised entirely of photometry from the \texttt{HSC Spec} sample.

\begin{figure}
    \centering
    \includegraphics[width=0.9\linewidth]{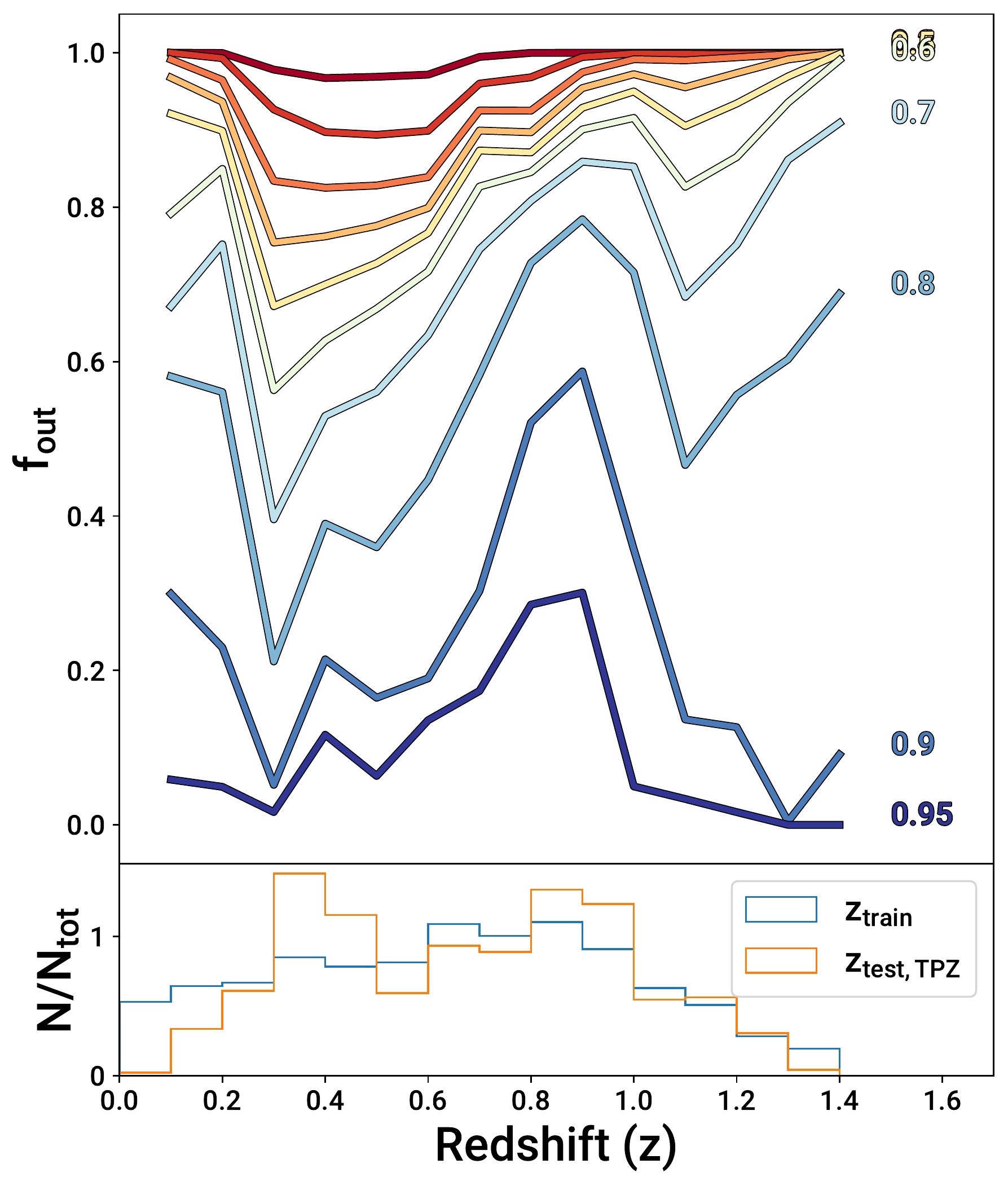}
    \caption{A plot of the sample fraction retained after various cuts in $C_\mathrm{NNC}$ as a function of redshift for the Match $\rightarrow$ COSMOS2015 Case detailed in Section \ref{Sec: MatchCOSMOS} (top) and the redshift distribution of the training (blue) and fitted test set (orange; bottom).  In the top panel, lines progress from red to blue with increasingly stringent cuts.  This plot shows that the NNC more quickly rules out galaxies at $z_\mathrm{phot}\sim 0.4$ and $z_\mathrm{phot}\sim1.2$ as having uncertain photo-$z$'s while a higher fraction of the sample is retained for galaxies at $z_\mathrm{phot}\sim0.2$ or $0.8 \lesssim z_\mathrm{phot} \lesssim 1$.  The NNC is able to select objects from across the full redshift range of the sample for most cuts in $C_\mathrm{NNC}$.}
    \label{fig:zfrac}
\end{figure}

    The fourth row of Figure \ref{fig:all_results} shows the results of this approach.  The left panel shows similar performance for the initial redshift fits to that of the Spec $\rightarrow$ COSMOS2015 Case (\S~\ref{Sec: SpecCOSMOS}).  Here, we find that the $zConf$ cut outperforms the the equivalent TPZ uncertainty cut resulting in a more accurate sample as well as improving the representation of galaxies at $z \gtrsim 1$ when selecting the best third of the test set.  However, the NNC-selected sample shown in the right panel produces similar sample representation at $z\gtrsim 1$ to that of the original photo-$z$ fit while also also providing improvements in outlier rate and standard deviation of $35\%$ and $23\%$ respectively in this most realistic case.  
    
    Using the results of Section \ref{Sec: COSMOSCOSMOS} as a best-case representation of model performance on \texttt{COSMOS2015} as a test set, we find that the outlier fraction increases from 0.036 to 0.055 and $\sigma_z$ increases from 0.046 to 0.062 for the Match $\rightarrow$ COSMOS2015 case.  The similarity in reported metrics between this case and the Spec $\rightarrow$ COSMOS2015 case indicates that while the matching process has increased the representation of dim galaxies in the training sample, in this particular case it is not able to supercede the performance of \texttt{HSC Spec} as a training sample.  Nonetheless, \added{while this analysis benefits from an expansive catalog of spectroscopic redshifts with a similar range in color-magnitude space to that of COSMOS2015, }the method of training set construction introduced here could be used to generate representative training sets for future surveys using a non-representative sample of spec-z's and a larger photometric catalog.

\section{Discussion} \label{Sec: Results}

The Match $\rightarrow$ COSMOS2015 case represents our most realistic  effort to design a training set that will provide good fits to the \texttt{HSC Phot} sample while using an independent test set to estimate fit accuracy.  As such, we will focus our discussion of results on this particular approach.  

Ultimately, our analysis aims to improve the selection of galaxy samples with accurate photo-$z$'s over a simple cut in uncertainty reported by a given photo-$z$ code.  To this end, we can plot the trade-off between sample accuracy and retained sample fraction by sorting galaxies either by TPZ-reported uncertainty or NNC confidence value, as shown in Figure \ref{fig:nfrac}.  The NMAD is nearly identical between the two, however the middle and right panels showing outlier fraction and standard deviation reveal that the NNC is greatly reducing the fraction of outliers.  In the middle panel in particular, we use a definition for outliers of $|\Delta z|/(1+z) > 0.15$.  This is a bit larger than the typical minimum tomographic bin width, but is consistent with other photo-$z$ literature \citep{Laigle2016, Tanaka2018} and thus represents outliers that cause significant cross-contamination with neighboring bins.  While we use Match $\rightarrow$ COSMOS2015 as our preferred case, these results look nearly identical for the Spec $\rightarrow$ COSMOS2015 Case.

For our preferred case of Match $\rightarrow$ COSMOS2015, we can cut the outlier fraction in half or more using our NNC compared to a naive TPZ uncertainty cut across a broad range of selected final sample sizes from 1-35\% of the original sample.  We find that the NNC reduces the outlier fraction and therefore $\sigma$ of the sample of galaxies relative to the TPZ uncertainty cuts.  It is important to note that there is a $\sim 7\%$ outlier rate inherent in the COSMOS2015 sample as noted in Figure \ref{fig:cosmos2015_zphot_szpec}.  However, the COSMOS2015 photometry extends an additional 2000$\mathrm{\AA}$ blueward and 9$\mu \mathrm{m}$ redward of the HSC $grizy$ bands in addition to greatly increased optical spectral resolution due to the addition of medium-band filters.  Because COSMOS2015 makes use of template-based photo-$z$ fitting code \textsc{LePhare} \citep{Arnouts2002, Ilbert2006} while TPZ is machine learning-based, many outliers produced by one of the codes should instead be fit well by the other.  As a result, \replaced{the  majority}{it is possible that many} of the outliers reported for the NNC selection shown in the last row of Figure \ref{fig:all_results} can be accounted for by the $\sim 5\%$ outlier rate of the \texttt{COSMOS2015} sample\added{, however it is impossible to tell the extent to which this is true without a more complete spectroscopic sample}.

When running a suite of photo-$z$ codes on similar HSC photometry, \cite{Tanaka2018} found that fits typically produced an NMAD of 0.05 and an outlier rate of 15\% (defined as $|\Delta z| > 0.15$) at $z \lesssim 0.2 \lesssim 1.5$.  As shown in Figure \ref{fig:nfrac}, we achieve an outlier fraction of a few percent at a similar sample accuracy using the NNC for the Match $\rightarrow$ COSMOS2015 Case.  Alternatively, COSMOS2015 claims an NMAD of 0.009 and outlier fraction of 0.0732 (defined as $|\Delta z|/(1+z) > 0.15$) for the FMOS sample of galaxies at $0.8 < z < 1.5$ \citep{Roseboom2012}.  We find, therefore, that the inclusion of the NNC in the photo-$z$ selection process effectively allows the selection of subsamples from \texttt{HSC Phot} with an improved outlier fraction and comparable NMAD to the COSMOS2015 30-band photo-$z$'s.

In order to investigate how the NNC affects the redshift distribution of the chosen galaxy sample, we plot the retained sample fraction as a function of redshift in the top panel of Figure \ref{fig:zfrac}.  This can be compared against the distribution of $z_\mathrm{train}$ and $z_\mathrm{test,TPZ}$ shown in the lower panel to see that this is not a case of over-emphasizing regions of redshift space that are over-represented in the training or test sets.  For example, only a small fraction of galaxies at $z\sim 0.4$ are selected with high confidence values even though there is an overabundance of galaxies at that redshift.  The variations in retained sample fraction shown in Figure \ref{fig:zfrac} can instead be interpreted as the reliability of the information provided by the $grizy$ bands at a particular redshift in determining galaxy redshift.

\begin{figure}
    \centering
    \includegraphics[width = 0.9\linewidth]{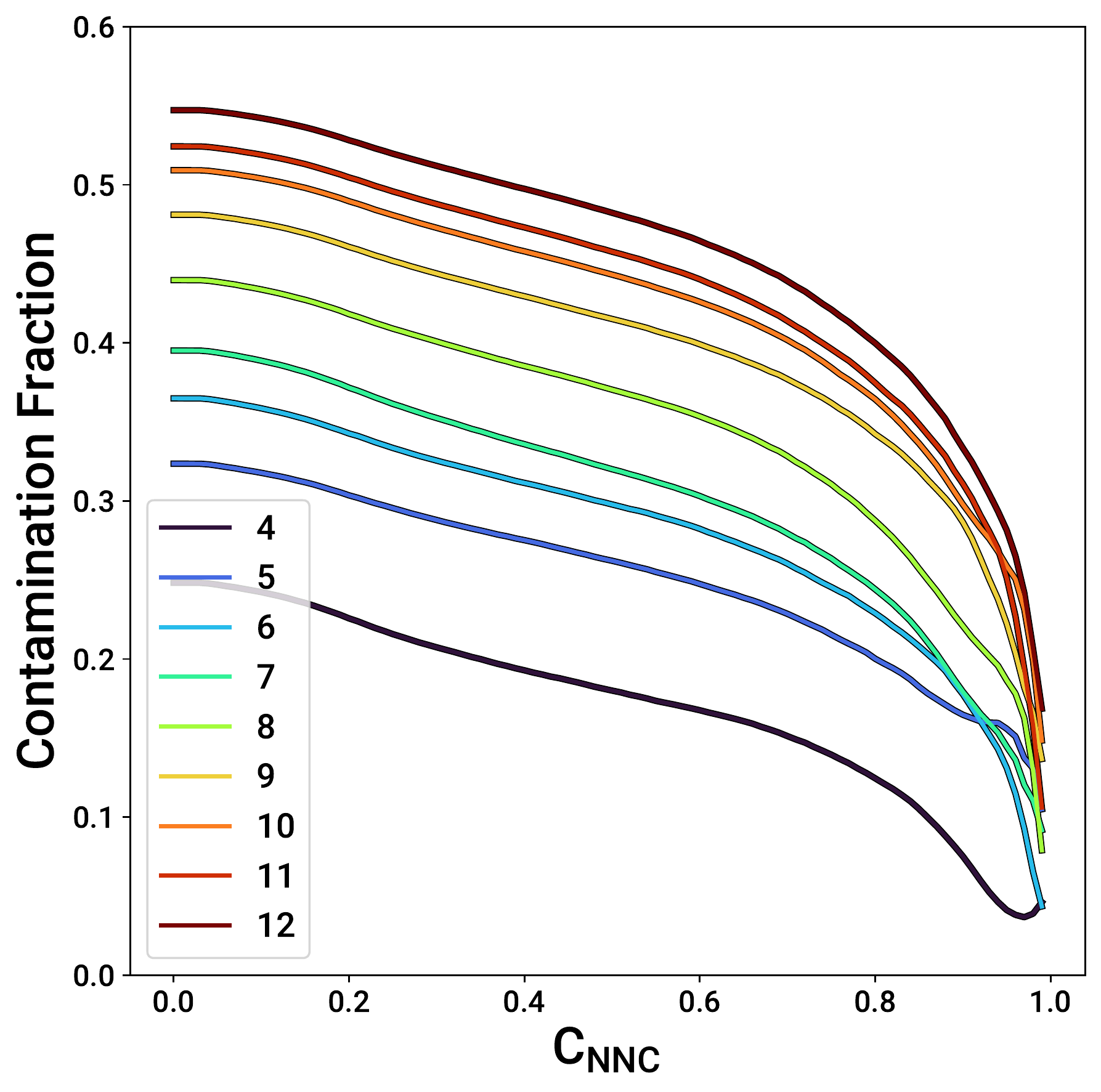}
    \caption{Sample contamination fraction (the number of objects incorrectly assigned to a redshift bin divided by the total number of objects) for various numbers of tomographic bins as a function of $C_\mathrm{NNC}$ cut.  The NNC is able to reduce the sample contamination fraction by $\sim0.1$ at $C_\mathrm{NNC}=0.8$ when compared to the default case of no selection ($C_\mathrm{NNC}=0.0$).}
    \label{fig:contam_frac}
\end{figure}

\begin{figure}
    \centering
    \includegraphics[width = 0.9\linewidth]{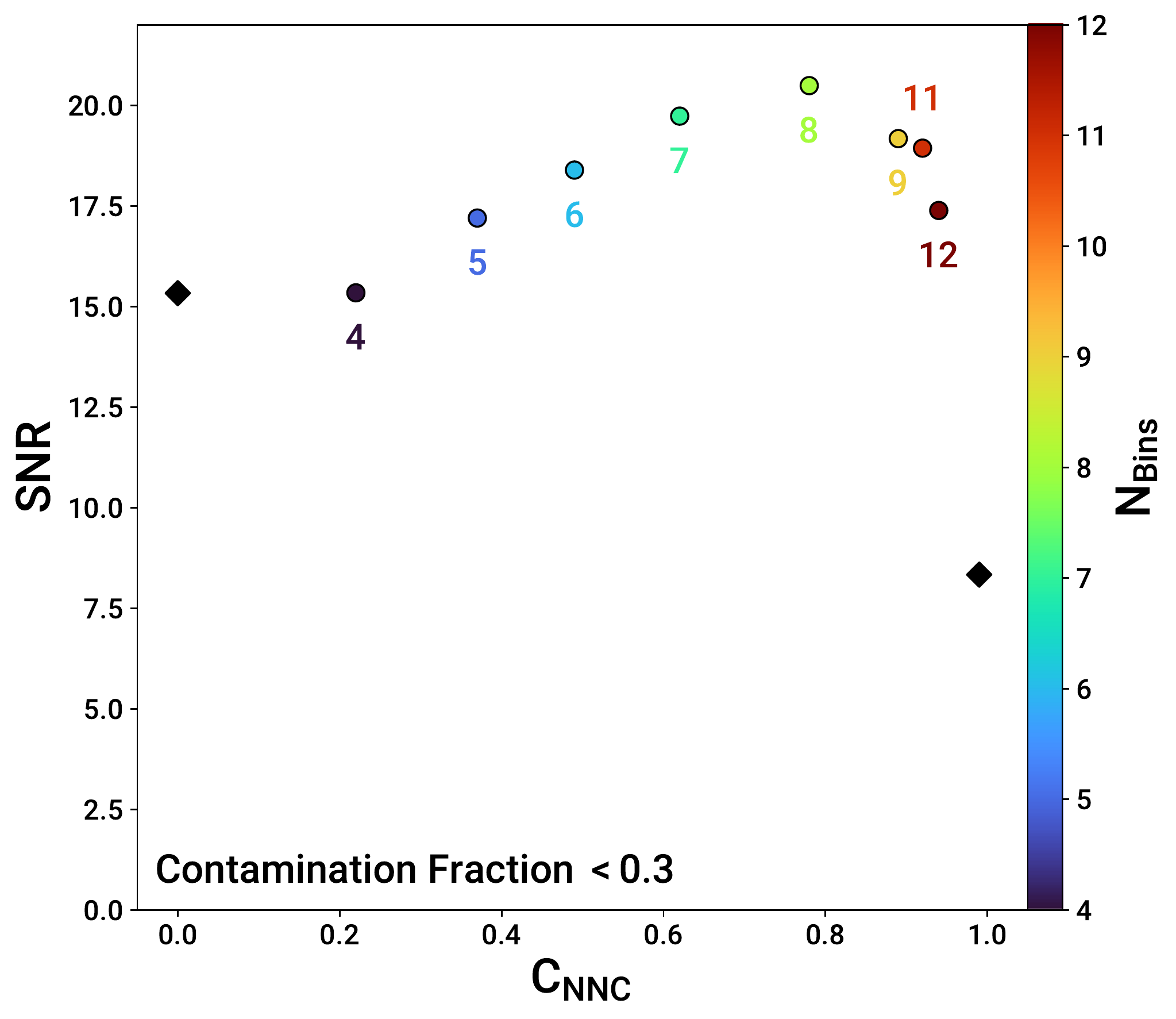}
    \caption{Total signal-to-noise ratio calculated from angular power spectra as a function of $C_\mathrm{NNC}$.  Each point is labeled with the number of tomographic bins used, which is chosen to be as large as possible while maintaining a contamination fraction of 0.3 or less.  Although a given number of bins is optimal over a range of $C_\mathrm{NNC}$ values, we simply plot the maximum point for each range for convenience.  Black diamonds mark the SNR calculated at $C_\mathrm{NNC} = 0.0$ and $C_\mathrm{NNC} = 0.99$ so the shape of the curve can be inferred for the extreme values of $C_\mathrm{NNC}$.  SNR is reduced at low $C_\mathrm{NNC}$ due to the low number of tomographic bins, while the decrease at high $C_\mathrm{NNC}$ is caused by reaching the shot noise limit when retaining $\sim1\%$ of the original galaxy sample.}
    \label{fig:snr_contam_frac}
\end{figure}

\added{To better understand the improvement in cosmological constraints from applying the NNC classifier, we compute the angular power spectra for the Match $\rightarrow$ COSMOS2015 sample for various cuts in $C_{NNC}$ and numbers of equal-sized bins covering the range $0<z<1.5$ using the Core Cosmology Library\footnote{\href{https://github.com/LSSTDESC/CCL}{https://github.com/LSSTDESC/CCL}} (CCL; \citealt{Chisari2019}).  We can stack all of the auto- and cross-power spectra into a data vector, $\mu (C_{\ell,ij})$ where $\ell$ is the wave number, and $i$ and $j$ are the redshift bins.  The Gaussian contribution to the covariance matrix of these two vectors can then be calculated by extending the derivation of \cite{Takada2009} to produce:
\begin{align}
    \mathbf{Cov}(C^\mathrm{obs}_{\ell,ij}, C^\mathrm{obs}_{\ell', i'j'}) = & \frac{\delta_{\ell\ell'}}{(2\ell + 1) \Delta \ell f_\mathrm{sky}} \nonumber \\
     & \times \left[ C^\mathrm{obs}_{\ell,ii'}C^\mathrm{obs}_{\ell,jj'} + C^\mathrm{obs}_{\ell,ij'}C^\mathrm{obs}_{\ell,ji'} \right]
\end{align}
where $\delta_{\ell\ell'}$ is the Kronecker delta function, $\Delta \ell$ is the width of $\ell$ bins, and $f_\mathrm{sky}$ is the sky fraction covered by the survey.  It is also important to note that $C^\mathrm{obs}_{\ell,ij} = C_{\ell,ij} + \delta_{ij}/\eta_i$ where $\eta_i$ is the average number density of galaxies in redshift bin $i$.  Finally, the overall signal-to-noise ratio is given by
\begin{align}
    \mathrm{SNR} = \mu^T \mathbf{Cov}^{-1} \mu.
\end{align}

It is generally desirable to split the sample into as many tomographic bins as possible to maximize SNR, but the choice of bin number must be weighed against shot noise and sample contamination.  The shot noise dominates the SNR calculation when there are too few galaxies per tomographic bin while large enough sample contamination can lead to significant biases in estimating the true redshift distribution of each bin.    We show the sample contamination as a function of $C_\mathrm{NNC}$ for different numbers of tomographic bins in Figure \ref{fig:contam_frac}, which shows an equal reduction in contamination fraction of $\sim 0.1$ over $0.0 < C_\mathrm{NNC} < 0.8$ regardless of the number of tomographic bins.  Further, Figure \ref{fig:snr_contam_frac} shows the calculated SNR when using the maximum possible number of bins while maintaining a contamination fraction of less than 0.3 (which we choose as a fiducial value).  The low-$C_\mathrm{NNC}$ region of the plot shows decreased SNR due to projection effects caused by using relatively wide photo-$z$ bins while $C_\mathrm{NNC} \gtrsim 0.9$ shows a sharp drop due to the reduction in galaxy number counts and a corresponding increase in the shot noise.  Meanwhile, a sample selection at $C_\mathrm{NNC} \sim 0.8$ maximizes the SNR by maintaining a low contamination fraction for 8 tomographic bins and shows a $\sim35\%$ increase in SNR over the original galaxy sample, which necessitates only 4 bins for a similar contamination fraction. }

\section{Conclusions}\label{Sec: Conclusions}

The LSST and similar future surveys will provide catalogs of billions of galaxies with measured photometry.  Photometric redshift fits will be performed on these catalogs, with the results informing galaxy studies of all kinds, including studies of galaxy physical properties and cosmological analyses.  To this end, we apply a custom Neural Network Classifier to photometric redshift fits produced by the public photo-$z$ fitting code TPZ in order to select galaxies with highly accurate redshfits.

We make use of the HSC DR2 Wide catalog due to its similarity to initial data releases planned for the LSST.  This includes a large photometric catalog as well as a smaller spectroscopic catalog, similar to what will be available from the LSST and DESI/4MOST spectroscopy.  Additionally, we use high-quality 30-band photometric redshifts from COSMOS2015 as an analogous data set to galaxies with superior photo-$z$ from the LSST Deep Drilling Fields.

We design four pairs of training and test sets for both TPZ and our NNC: the Spec $\rightarrow$ Spec,  Spec $\rightarrow$ COSMOS2015, COSMOS2015 $\rightarrow$ COSMOS2015, and Match $\rightarrow$ COSMOS2015.  These are designed to make the training set increasingly representative of the \texttt{HSC Phot} sample, which is our nominal application set, with the second and fourth cases estimating realistic performance for test sets that are independent of the training set.    

The Spec $\rightarrow$ Spec Case, though highly optimistic in offering a fully representative training set with high S/N photometry, is helpful to see the ideal case of TPZ's photo-$z$ performance, as well as the NNC's performance on a sample that already has reliable photo-$z$'s.  We find that even in the event of a fully representative training set, the NNC is able to improve the standard deviation in $\Delta z/(1+z)$ ($\sigma_z$) and outlier fraction ($f_\mathrm{out}$; defined as the sample fraction with $|\Delta z|/(1+z) > 0.15$) over similar cuts made in $\sigma_\mathrm{TPZ}$ and $zConf$. The Spec $\rightarrow$ COSMOS2015 Case models the expected performance of the pipeline in the standard approach of training on \texttt{HSC Spec} and fitting \texttt{HSC Phot}.  Here, the NNC provides a modest improvement in outlier fraction and $\sigma_z$, but noisy photo-$z$ fits caused by the mismatch between the training and test sets persist even after the NNC selection.  The COSMOS2015 $\rightarrow$ COSMOS2015 Case provides a best-case estimate for the pipeline's performance using the intersection of the \texttt{HSC Phot} and \texttt{COSMOS2015} samples as a test set and, as expected, showed some degradation from the Spec $\rightarrow$ Spec Case.  Finally, the Match $\rightarrow$ COSMOS2015 Case implements data augmentation to construct a training sample that should better represent  the \texttt{COSMOS2015} sample before evaluating the NNC performance on COSMOS2015 30-band photo-$z$'s.  This results in similar photo-$z$ fits to the Spec $\rightarrow$ COSMOS2015 Case, with a similar improvement in photo-$z$ accuracy after the NNC selection is performed.  

Selecting galaxies using the NNC confidence, $C_\mathrm{NNC}$, outperforms a naive cut in TPZ's reported uncertainty (or photo-$z$ Gaussianity measure $zConf$), across almost all possible retained sample fractions.  In particular, for our most sophisticated representation of training and test sets, the Match $\rightarrow$ COSMOS2015 Case, we find a 35\% reduction in outlier fraction and a 23\% improvement in standard deviation of the photo-$z$ for the example case of selecting the most accurate third of the test set.  The NNC is able to match or outperform selections made using $\sigma_\mathrm{TPZ}$ or $zConf$ for any selected sample size in terms of photo-$z$ standard deviation and outlier fraction, while for small retained samples it produces a slight degradation in NMAD compared to those approaches.  
 
Additionally, we find that the NNC is able to select galaxies from a broad range of redshifts for inclusion in the final sample.  For surveys like the LSST, this will enable a pipeline like ours to select smaller galaxy samples with significantly reduced outlier fractions and hence higher S/N for cosmological analysis.  The non-uniform selection in redshift requires updating the predicted redshift distribution, N(z), and it may be preferable to update the choices of bin edges to obtain sufficient statistics to remain above the shot noise limit.  We find that selecting the most accurate third of the galaxy photo-$z$ sample using any of the methods in this work yields a significant improvement in outlier fraction, NMAD, and $\sigma_z$.  In the most realistic case, using the NNC to select a third of the original galaxy sample achieves a factor of two reduction in photo-$z$ standard deviation and a factor of five reduction in outlier rate.   
 
\added{We use CCL to calculate angular power spectra for the Match $\rightarrow$ COSMOS2015 Case using various $C_\mathrm{NNC}$ selections and number of tomographic bins.  We demonstrate that cuts in $C_\mathrm{NNC}$ are capable of reducing the contamination fraction by 0.1 for virtually any number of tomographic bins.  Further, when choosing the maximum number of tomographic bins such that the contamination fraction is below 0.3, we find that selections made with $C_\mathrm{NNC}$ increase the measured SNR by $\sim35\%$ by enabling an analysis with 8 tomographic bins rather than the 4 bins necessary for the unselected sample.  A full implementation of the method illustrating the resulting improvement in cosmological parameter estimation is an important next step and is planned as a follow-up investigation.}

\section*{Acknowledgements}
The authors would like to thank Sam Schmidt for his expert advice on running BPZ and TPZ, as well as the insightful comments and suggestions from Viviana Acquaviva, Humna Awan, Kartheik Iyer, and John Wu.  \added{The authors would also like to thank the referee, whose perceptive suggestions significantly improved the quality of this analysis during the review process.} This material is based upon work supported by the U.S. Department of Energy, Office of Science, Office of High Energy Physics Cosmic Frontier Research program  under  Award \replaced{Numbers  DE-SC0011636  and }{Number} DE-SC0010008
%which supported AB and EG
. 

\appendix

\section{Training Sample Size Verification}

\begin{figure}
    \centering
    \includegraphics[width = 0.8\textwidth]{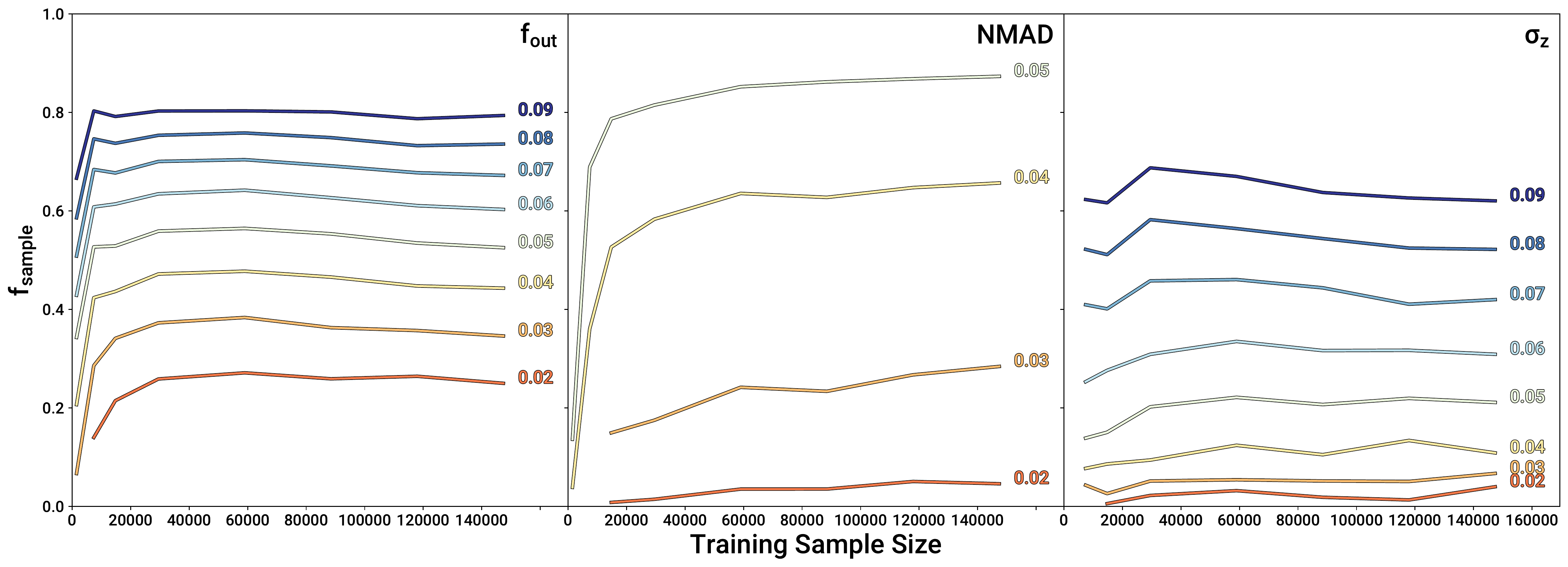}
    \caption{In a similar layout to Figure \ref{fig:acc_var}, we plot retained sample fraction as a function of training sample size for various choices of outlier fraction, NMAD, and $\sigma_z$.  While $f_\mathrm{out}$ and $\sigma_z$ show no improvement with increased training set size after $\sim50\,000$ objects, the NMAD shows monotonic (though small) improvement up to and including the full training sample.  This indicates that increasing the training sample size would continue to improve the core of the photo-$z$ error distribution, but that it would not have an effect on outlier reduction.}
    \label{fig:trainsize_var}
\end{figure}

We show that the performance of the NNC in the Match $\rightarrow$ COSMOS2015 case is not limited by the training sample size in Figure \ref{fig:trainsize_var}.  Plotting the training sample size against the standard deviation of the photo-$z$ fit error for 1\%, 5\%, 10\%, 20\%, 40\%, 60\%, 80\%, and 100\% of the full training sample, we find that training on $\sim 50\,000$ objects (which is roughly 30\% of the available training sample in this case) is sufficient to maximize the NNC selection performance for outlier rejection in the Match $\rightarrow$ COSMOS2015 Case.  The NMAD shows incremental, but monotonic improvement with increasing sample size, suggesting that a large training sample could continue to yield small improvements to the width of the photo-$z$ error distribution core.

\section{Results When Using BPZ Photo-z Fits} \label{apdx:bpz}
\begin{figure*}
    \centering
    \includegraphics[width = \textwidth]{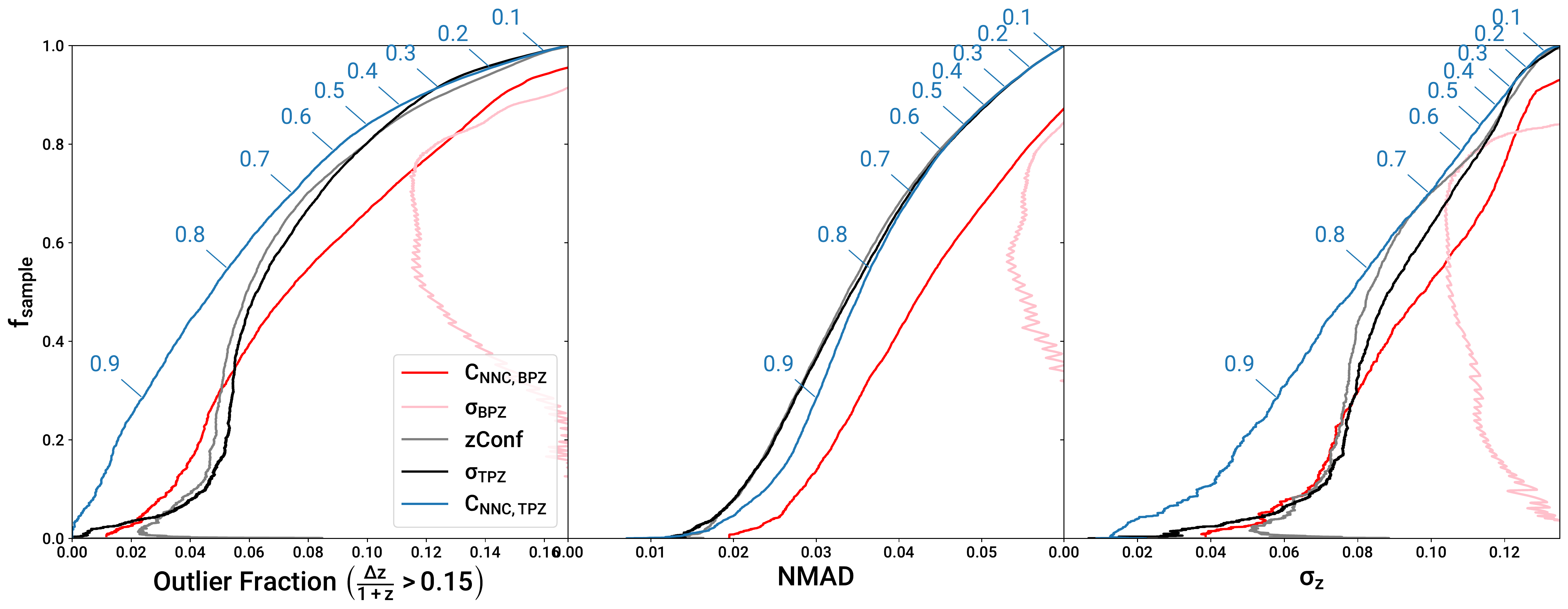}
    \caption{Similar results to Figure \ref{fig:nfrac} with the addition of curves for cuts in reported BPZ uncertainty ($\sigma_\mathrm{BPZ}$;light red) and an NNC trained on the BPZ redshifts ($C_\mathrm{NNC,BPZ}$; dark red).  Selections using $\sigma_\mathrm{BPZ}$ are unable to improve sample accuracy after rejecting more than $\sim30\%$ of the sample, while $C_\mathrm{NNC,BPZ}$ gets similar results to $\sigma_\mathrm{TPZ}$ and $zConf$ in outlier fraction (middle) and photo-$z$ error standard deviation (right).  It is unable to match the NMAD of any selector from the TPZ-based pipeline, with a roughly constant increase of $\sim 0.008$ across all retained sample fractions.}
    \label{fig:nfrac_bpz}
\end{figure*}

Throughout this work, we make use of TPZ to provide initial photometric redshift fits, however our NNC afterburner can be applied to any photo-$z$ code.  As an example, Figure \ref{fig:nfrac_bpz} shows the results of Figure \ref{fig:nfrac} with additional curves in red for Bayesian Photo-Z (BPZ).  This figure shows the decreased overall accuracy provided by BPZ fits, as indicated by the higher NMAD, outlier fraction, and $\sigma_z$ values at $f_\mathrm{sample}=1$ for the red curves.  Additionally, the light red curves indicate that, while limiting the sample based on reported BPZ uncertainty initially improves sample photo-$z$ accuracy, decreasing the sample by more than a factor of $\sim2$ results in a significant decrease in accuracy.  The dark red curve, on the other hand, indicates that the NNC is able to achieve an outlier fraction and photo-$z$ error standard deviation similar to that of the pre-NNC TPZ fits, with a constant increase in the NMAD of $\sim 0.008$ for all retained sample fractions.  As our interest was simply to investigate the efficacy of the NNC afterburner on a second photo-$z$ code, we have not tried to optimize the BPZ performance for this dataset.

\section{Implementation of a Neural Network Regressor Prior to Classification}

One of the tools we employed in an effort to further improve sample selection and photo-$z$ accuracy was an intermediate stage implementing a neural network regressor (NNR).  The structure of the regressor is identical to that of the NNC (i.e. four hidden layers with [100, 200, 100, 50] neurons and a single output neuron) with the exception that the output neuron was assigned a linear activation function.  The NNR is trained to regress for $z_\mathrm{phot} - z_\mathrm{spec}$ using TPZ outputs ($z_\mathrm{phot}$, $\sigma_\mathrm{TPZ}$, $zConf$) and photometry ($i$-band magnitude, colors, band triplets) such that a corrected redshift can be produced: $z_\mathrm{NNR}$.  In this way, the function of the NNR is to correct systematic errors inherent in the particular photo-$z$ code that produced the initial fits (which is TPZ for our case).

With the addition of the NNR to correct for photo-$z$ errors, we also modified the NNC to add $z_\mathrm{NNR}$ as a feature in addition to produce two outputs rather than one.  Each output produces an independent confidence value from 0 to 1 indicating whether $z_\mathrm{NNR}$ or $z_\mathrm{TPZ}$ is within $|\Delta z|/(1+z_\mathrm{spec}) < 0.10$ respectively.  The results could then be consolidated into a single point redshift estimate and confidence value by choosing the higher confidence value and its corresponding redshift.  With the current dataset, the NNR did not lead to a significant improvement and was dropped from the pipeline, but this step is worth considering for future applications of our approach.  

\begin{figure}
    \centering
    \includegraphics[width = 0.8\textwidth]{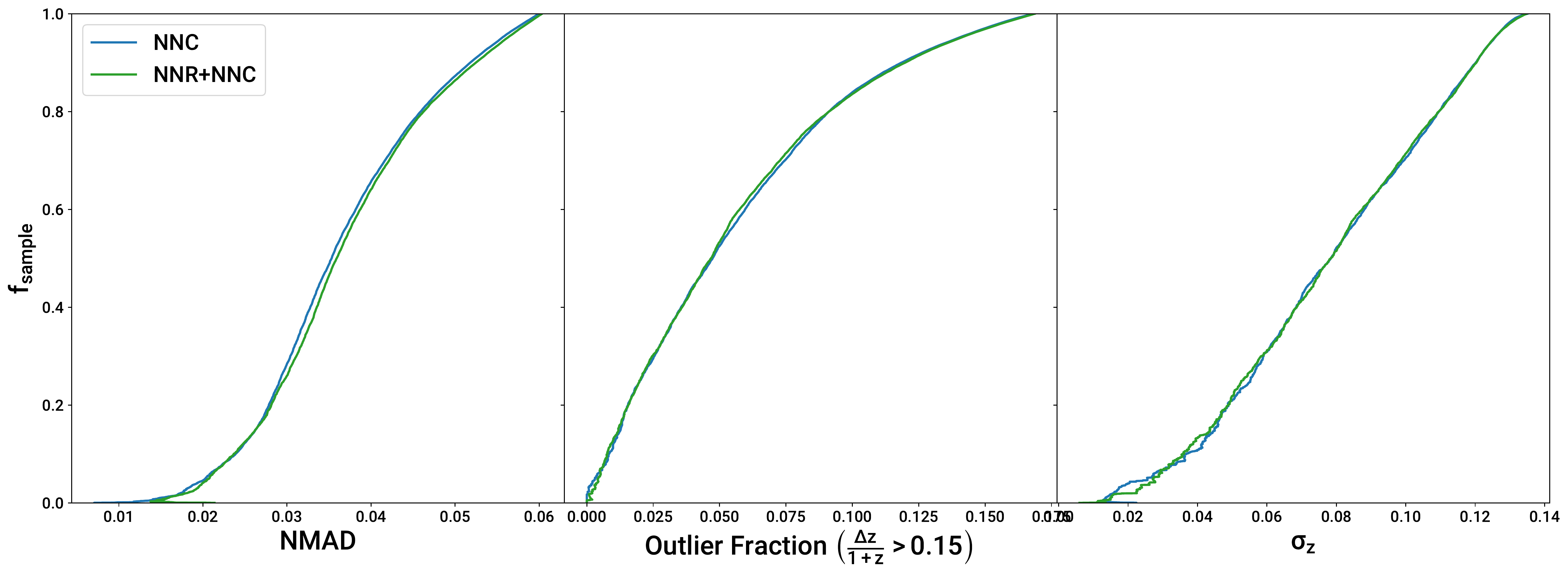}
    \caption{The three panels are similar to those of Figure \ref{fig:triplet_comparison}, however here we show a comparison between selecting galaxies using a NNC alone (blue) vs. using a NNR to regress for TPZ's photo-$z$ error and selecting galaxies using the corrected redshifts for the Match $\rightarrow$ COSMOS2015 Case.  All metrics of accuracy are nearly identical for all possible sample fractions, indicating that the addition of the NNR to the pipeline does not provide meaningful improvements to the photo-$z$ selection pipeline.}
    \label{fig:nnr_comparison}
\end{figure}

\bibliographystyle{apj_url}
\bibliography{bibliography}

\end{document}